# Scale-Free Distribution of Oxygen Interstitial Wires in Optimum-Doped HgBa$_2$CuO$_{4+y}$

Gaetano Campi [1,*], Maria Vittoria Mazziotti [2,3], Thomas Jarlborg [2,4] and Antonio Bianconi [1,2,*]

[1] Institute of Crystallography, CNR, Via Salaria Km 29.300, 00015 Monterotondo, Italy
[2] Rome International Center for Materials Science Superstripes RICMASS, Via dei Sabelli 119A,
00185 Roma, Italy; vittoria.mazziotti@gmail.com (M.V.M.); thomas.jalborg@unige.ch (T.J.)
[3] Department of Mathematics and Physics, University Roma Tre, Via della Vasca Navale 84,
00146 Roma, Italy
[4] DQMP, University of Geneva, 24 Quai Ernest-Ansermet, CH-1211 Geneva, Switzerland

**Abstract:** Novel nanoscale probes are opening new venues for understanding unconventional electronic and magnetic functionalities driven by multiscale lattice complexity in doped high temperature superconducting perovskites. In this work we focus on the multiscale texture at supramolecular level of atomic oxygen interstitials (O-i) stripes in HgBa$_2$CuO$_{4+y}$ at optimal doping for the highest superconducting critical temperature $T_C$=94K. We report compelling evidence for the nematic phase of oxygen-interstitial O-i atomic wires with fractal-like spatial distribution over multiple scales by using scanning micro and nano X-ray-diffraction. The scale free distribution of O-i atomic wires at optimum doping extending from micron scale down to nanoscale has been associated with the intricate filamentary network of hole rich metallic wires in the CuO$_2$ plane. The observed critical opalescence provides evidence for the proximity to a critical point controlling the emergence of high temperature superconductivity at optimum doping.

## 1. Introduction

The true nature and the key role of the complexity at nanoscale in high-temperature superconductors [1–8] and related systems [9] have sparked growing interest in the condensed matter research field, since the discovery of unconventional overdoped perovskites [10,11]. The development of novel synchrotron radiation sources and imaging techniques on the basis of focusing on the X-ray beam down to the nanoscale allow for the visualization of multiscale inhomogeneities of the supramolecular structure [12–15]. In this way, scale invariant textures have been observed in lattice, spin, and charge degrees of freedom in strongly correlated oxide perovskites [16–21] and at metal-insulator transitions [22]. These results have shown that scale invariance [23,24] in granular matter near a critical point has been favoring quantum coherence in high-temperature superconductivity [25–29]. Lattice complexity beyond the average crystalline structure effects in the mechanism of high $T_c$ superconductivity at optimum doping is attracting high interest. In particular, the role of oxygen interstitials and vacancies spatial disposition has a strong impact on the electronic properties of superconducting perovskite materials [30–32]. The texture of oxygen interstitials (O-i) related with their self-organization due to their large mobility in the space layers above about 200 K was detected by advanced scanning micro-X-ray diffraction, in layered perovskites, such as nickelates $La_2NiO_{4+y}$ [33,34] and in several families of cuprate high-temperature superconductors (HTS), such as doped $La_2CuO_{4+y}$ (La214) [35–42], $YBa_2Cu_3O_{6+y}$ (Y123) [43–45], and $Bi_2Sr_2CaCu_2O_{8+y}$ (Bi2212) [46–49]. In this work, we focus on $HgBa_2CuO_{4+y}$ (Hg1201) at optimum doping [50–60], which is the simplest member of the mercury-based cuprate compounds that provides the highest superconducting transition temperature. $HgBa_2CuO_{4+y}$ has a simple tetragonal average structure with average Cu-O bond length of 194 pm [54], showing self-organization of a nematic phase with co-existing O-i atomic stripes [56,57] running in the a-direction (100) and b-direction (010) of the ab plane. The presence of anisotropic structures at room temperature, and the understanding of their properties, are crucial to explain the still elusive room temperature nematicity observed in HTS [61,62]. However, while the nanoscale phase separation made of first puddles rich in O-i stripes anticorrelated with second puddles that show short-range charge density waves (CDW) has been visualized by scanning micro-X-ray diffraction [57], the imaging of the nanoscale phase separation on the mesoscale (100–1000 nm) in Hg1201 is still missing. Information on this intermediate scale appears to be quite important, since the oxygen interstitials form domains with a dimension ranging from few, up to tens of nanometers. Therefore, higher resolution in real space is required for the X-ray diffraction measurements to image oxygen interstitials ordering on the mesoscale. Thanks to the advances in X-ray, focusing on optics is nowadays possible to cover this length scale connecting the microscopic to the nano- and atomic-world using beams down to 100 nm in scanning mode.

In this work, we have used scanning micro-X-ray diffraction (SµXRD) with a focused beam of 1 × 1 µm² joint with scanning nano-X-ray diffraction (SnXRD) with a focused beam of 100 × 300 nm$^2$ for directly imaging the texture formed by the oxygen interstitials in Hg1201 at microscale and mesoscale. We exploited spatial statistic tools, such as the probability density function of O-i populations, spatial correlations, and connectivity to describe quantitatively the O-i textures from micron to mesoscale. The results confirm the power law behavior of the oxygen interstitials organized in stripes at micrometric scale in Hg1201 [57], which was observed first in La$_2$CuO$_{4+y}$ at optimum doping [35,39]. In addition, at improved resolution, we have been able to visualize a percolation interface between the insulating CDW puddles and the oxygen wires [61,62].

## 2. Results

The HgBa$_2$CuO$_{4+y}$ single crystals have been grown with a final oxygen treatment to establish the y concentration of oxygen interstitials of about y = 0.12 showing the superconducting optimum critical temperature (T$_C$) of 94 K [51–53]. The crystal structure has been determined by standard X-ray diffraction. The structure, schematized in Figure 1a, has a tetragonal P4/mmm space group symmetry, and the unit cell parameters are: a = b = 3.874 Å, c = 9.504 Å at room temperature in agreement with reference [54].

Nematic phase of atomic oxygen wires has been investigated by high-energy X-ray diffraction measurements performed at the BW5 beamline of the DESY Synchrotron in Hamburg. The results provide compelling experimental evidence for the location of the oxygen interstitial (O-i) content, y, which occupies the interstitial site ½, ½, and 0 in the basal plane of HgBa$_2$CuO$_{4+y}$. The arrangement of O-i gives rise to intriguing diffraction features constituted by clear diffuse streaks crossing the Bragg peaks along all the three crystallographic directions [56–58]. In Figure 1b, we show a typical diffuse streak between the (3,3) and (3,4) reflections in the hk diffraction plane. Figure 1c shows the streak profile in red and the background in black. The normalized X-ray diffraction profile is shown in the lower panel, where the diffuse streak intensity has been quantified by the difference between the Bragg peaks profile and the background. In this way, we obtained the streak intensity, due to the O-i stripes. The connection of the observed streaks with the O-i arrangement in stripes has been proved by calculating the X-ray diffraction pattern of a HgBa$_2$CuO$_{4.12}$ lattice made by 50 × 50 × 20 unit cells. In this model structure, the O-i located with positive correlations along the a- and b-axes has been created by the Monte Carlo method [63] using the software package DISCUS [64].

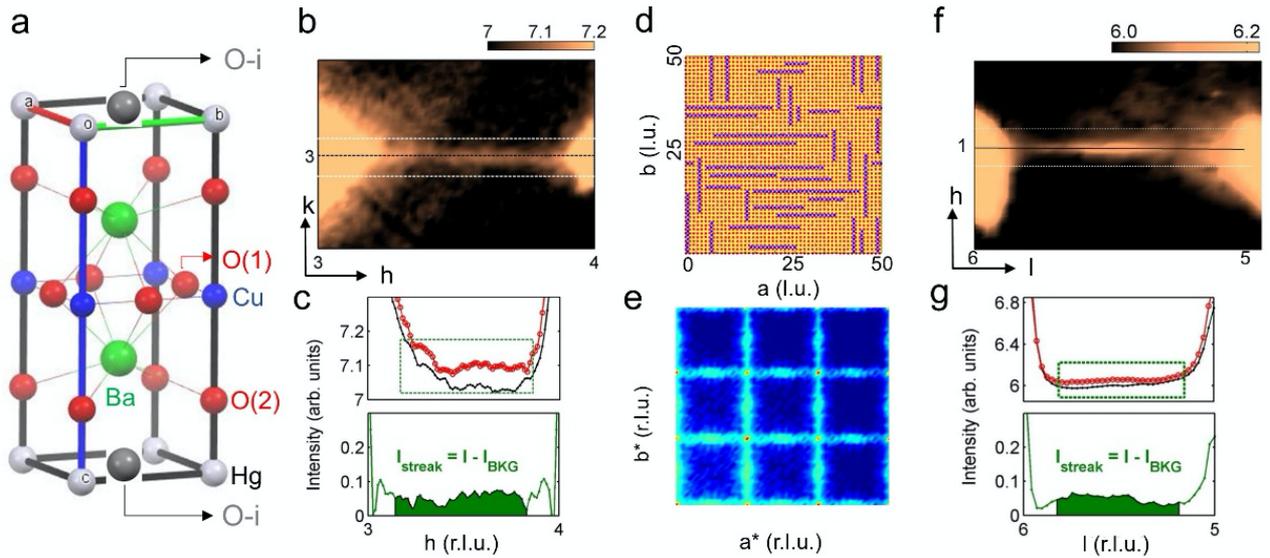

**Figure 1.** (**a**) Schematic representation of the crystal structure of tetragonal HgBa$_2$CuO$_{4+y}$. The oxygen sites O(1), O(2) and the interstitial O-i are shown. (**b**) A typical streak in the hk plane of the reciprocal space connecting two principal XRD reflections of tetragonal Hg1201 single crystal. (**c**) The diffraction intensity (red circles) along the h direction indicated by the black horizontal line and the background (black points) obtained by merging profiles along (h,3 ± ε) directions indicated by the white dashed lines in (**b**). The difference between the diffraction intensity and the background, calculated in the h range indicated by the dotted rectangle in (**b**), giving the streak intensity value, I$_{streak}$ (green area) is shown in the lower panel. (**d**) Structural model of ordered O-i in a portion of crystal with 50 × 50 × 20 unit cells. The O-i (violet spheres) are placed in unidimensional chains by the Monte Carlo method considering a positive correlation of 0.5 along the a and b directions with a concentration of 12%. (**e**) Fourier transform of the generated structure for a lattice composed by 50 × 50 × 20 cells in the hk plane of X-ray transmission pattern. We can observe diffuse streaks along the a* and b* directions, in qualitative agreement with diffraction measurements. (**f**) O-i diffuse streak in the hl plane of the reciprocal space connecting the (1, 5) and (1, 6) XRD reflections of Hg1201. (**g**) Diffraction profile (red circles) along the l direction and the background (black points) obtained by merging profiles along (1±ε, l) directions indicated by the white dashed lines in (**f**). The difference between the diffraction profile and the background, calculated in the l range indicated by the dotted rectangle in (**a**), giving the streak intensity value, I$_{streak}$ (green area) is shown in the lower panel.

We obtained qualitative agreement with the experimental diffraction pattern, generating a model of crystal structure with O-i on the ½, ½, and 0 positions, forming stripes along both the a and b directions. Figure 1d shows a pictorial view of O-i arrangement along the a and b crystallographic directions. In Figure 1e, the Fourier transform of the modeled structure is shown and it presents the streaks along the a* and b* directions, in agreement with the measured pattern shown in Figure 1b. This demonstrates that the observed diffuse streaks are determined by the O-i stripes, where the average concentration of O-i is 12% over a lattice of 50 × 50 × 20 unit cells corresponding to about 20 × 20 × 20 nm$^3$.

Scale-free distribution of oxygen wires from microscale to mesoscale has been probed first by scanning micro XRD measurements (SµXRD) and thus scanning nano XRD measurements (SnXRD) to enhance the spatial resolution. Both SµXRD and SnXRD measurements have been carried out at the ID13 beamline of ESRF in Grenoble. The O-i stripes in Hg1201 run along the a(b) direction with no correlation along the c direction; therefore, they give streaks also in the hl plane. The streak intensity, $I_{streak}$, in the hl plane has been calculated for each frame collected in the reflection geometry during the scanning of the sample, similar to what has been carried out for the $I_{streak}$ calculations in the hk plane (see Figure 1f,g).

In this way, we have built the maps of the streak intensity, $I_{streak}$, that visualize the O-i spatial distribution. The maps measured in SµXRD and SnXRD show rich stripe-ordered O-i (yellow regions) embedded in a matrix of disordered O-i (black regions), as shown in Figure 2a,b, respectively. To characterize this inhomogeneity, we used a spatial statistics approach. More specifically, we have computed the probability density function (PDF) spatial correlations, G(r), and percolation pathways of the streak intensity for both SµXRD and SnXRD. The PDF can be modelled by a power law behavior given by $PDF(I_{streak}) = C(I_{streak})^{-\alpha}$, where C is a constant. We find the critical exponent $\alpha = 2.0 \pm 0.1$ at both micron and mesoscale; the $I_{streak}$ of SnXRD has been scaled in a way that its PDF overlaps the PDF of $I_{streak}$ in the SµXRD, as shown in Figure 2c. The intensity of streak values ranges from values smaller than 1 in the poor oxygen wire regions to values larger than $10^3$ in the rich O-i wire zones. The results show the scale-free organization of O-i in the sample, as it was already found at microscale in other oxygen-doped cuprates. However, we observe that in the SµXRD maps, the size of oxygen wire domains does not exceed the single pixel size corresponding to an area of 1 × 1 µm$^2$ given by the scanning beam size. This limitation has been overcome by enhancing the real space resolution in XRD setup using SnXRD, where the beam size has been reduced to 0.1 × 0.3 µm$^2$. Indeed, in this case, the submicron structural features of domains rich in oxygen atomic wires can be appreciated (see Figure 2b).

To quantify the submicron structural features, we have extracted the spatial correlation function G(r) shown in Figure 2d. We observe that the decay of G(r) corresponds to 2 µm in both SµXRD and SnXRD, indicating this distance as the typical size of domains rich in oxygen atomic wires.

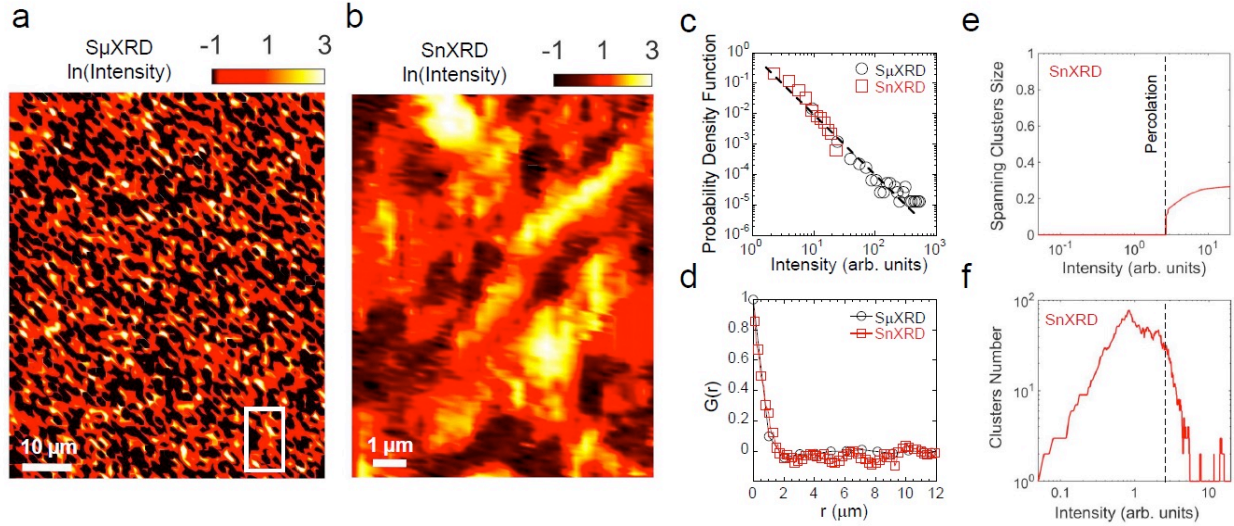

**Figure 2.** (**a**) Logarithmic spatial map of streak intensity extracted by SµXRD with 1 × 1 µm$^2$ spatial resolution. The rich stripes-ordered O-i (yellow regions) are anticorrelated with CDW rich domains (black regions). The bar corresponds to 10 µm. (**b**) SnXRD spatial map measured with 0.1 × 0.3 µm$^2$ of streak intensity in logarithmic scale. Here, we can appreciate the interface (red regions) between the rich stripes-ordered O-i (yellow regions) and CDW puddles (black regions). The bar corresponds to 1 µm and the size of the map corresponds to the size of the white rectangle in (**a**). (**c**) Plot of the probability density function of rich stripes-ordered zones relative to the SµXRD (open squares) and SnXRD (full circles) maps in (**a**,**b**). Both distributions have a power law exponent α = 2. (**d**) G(r) calculated on the SµXRD (open squares) and SnXRD (full circles) maps in (**a**,**b**). (**e**) Spanning clusters size and (**f**) forming clusters number calculated for the maps in (**b**). The percolation thresholds are represented by the vertical dashed lines.

We have studied the connectivity between the domains rich in O-i wires using cluster analysis. More specifically, we have calculated all the forming clusters, selecting the cluster with the largest extent, as a function of the streak intensity, $I_{streak}$. When this extent is equal to the system size, there is a spanning cluster and the system percolates. Therefore, we have calculated the percolation threshold, $P(I_{streak})$, the spanning cluster size, and the number of forming clusters in both SµXRD and SnXRD maps in Figure 2a,b. The results are shown in Figure 2e,f. At micron resolution, we cannot find a percolation threshold due to the resolution limit. On the other hand, in SnXRD map, we can find small clusters forming up to the percolation threshold, $P(I_{streak}) = 2.7$. At this intensity value, a spanning cluster occurs as an intermediate space (red color in the maps of panels a and b) connecting zones rich in oxygen wires (yellow spots in the maps) and CDW puddles (black regions).

Band structure calculations of doped HgBa$_2$CuO$_{4+y}$ at optimum doping with 0.12 < y < 0.16 have been carried out with the assumption of very large 1D and 2D superstructures of ordered O-i [60]. The band structure calculations agree on the strong inhomogeneity of hole doping in metallic CuO$_2$ plane near and far from the O-i oxygen interstitial wire. Following the experimental atomic inhomogeneity of doped HgBa$_2$CuO$_{4+y}$ crystal, its structure is formed by a nematic phase of O-i rich metallic wires made of HgBa$_2$CuO(5) unit cells [CuO(1)$_2$]*[BaO(2)]*[HgO(3)]*[BaO(2)]* separated by undoped portions running in the 100 and 010 directions, made of HgBa$_2$CuO$_4$, tetragonal units [CuO(1)$_2$][BaO(2)][Hg][BaO(2)]. The O-i sites in HgBa$_2$CuO(5) occupy a rather open empty space, which normally is unoccupied in HgBa$_2$Cu$_4$. Possible lattice deformations around O-i impurities are not considered in the calculations [61]. The charge density is strongly inhomogeneous and differs considerably between sites in the proximity and far away from the wire of oxygen interstitials.

Figure 3 shows the local decomposition of the DOS at $E_F$ (in units of (atom eV)$^{-1}$) for Hg$_{12}$Ba$_{24}$Cu$_{12}$O$_{48}$ and "striped" Hg$_{12}$Ba$_{24}$Cu$_{12}$O$_{48+n}$ with n = 2 oxygen interstitials. Apical O2 and planar O1 are the apical Oap and planar Opl oxygen coordinated by Cu. A periodic boundary condition was considered as an artificial periodicity of 6-unit cells. The supercell Hg$_{12}$Ba$_{24}$Cu$_{12}$O$_{48+2}$ extends over 6- and 2- lattice constants along x and y, respectively, and two additional oxygen interstitials were inserted at O(3) site in the Hg-plane forming an atomic stripe running along y. At a distance of 3-unit cells from the oxygen interstitial wire, the local electronic structure becomes close to the undoped system. Panel 3a shows the total density of states (DOS) for superconducting HgBa$_2$CuO$_{4+y}$ with the one-dimensional oxygen interstitials O-i in the [HgO-i] spacer layers and for the undoped cuprate. Figure 3b shows the partial DOS in the apical oxygen sites and Figure 3c the partial DOS in the atomic Hg plane.

The density of states at the Fermi energy, N($E_F$), is higher than the undoped system, about 3 eV$^{-1}$ per elementary cell. The increase in DOS is limited to the Cu adjacent to the O-i wire; see Figure 3 and Table 1. Ba and Hg atoms are negligible N($E_F$) in the undoped case, but the striped O-rich system shows a large increase in DOS on the sites near oxygen interstitials [60]. Hybridization with the p-states on the oxygen interstitials creates a large s- and d-DOS on Hg, and increase in the p-DOS on apical and planar oxygen states. The increase in the local DOS can be understood as an effect due to the O-i atomic stripe, which forms electronic wires with high hole doping when $E_F$ is approaching the wire van Hove DOS peak. The local DOS is very small in more distant atoms in the other atomic layers (BaO, Hg) with values very similar to the undoped Hg1201. At two lattice units from the O-stripe, the DOS already seems to have restored its local character as an undoped Hg1201, except for the modest increase in Cu-d DOS. Moreover, this is corroborated by the electric charge density within the Cu atoms, as well as by an analysis of the folded FS [60].

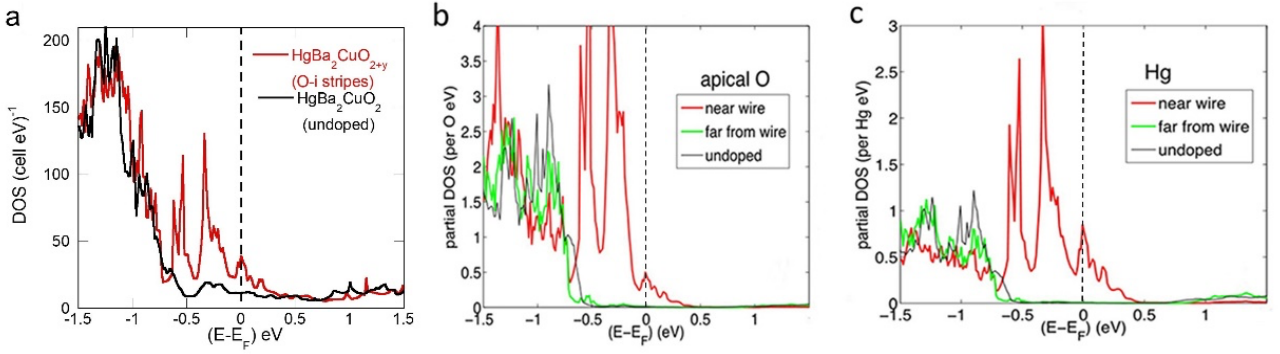

**Figure 3.** (**a**) Total density of states (DOS) for superconducting $HgBa_2CuO_{4+y}$ doped by oxygen interstitials O-i, forming one-dimensional oxygen interstitial O-i wires in the $HgO_y$ layers (red curve) showing a DOS peak at the Fermi level ($E_F$) due to a van Hove singularity (VHS). Panels (**b**,**c**) show the partial DOS near and far from the O-i stripes on the apical oxygen sites and on the atomic [Hg] plane (hosting O-i stripes), respectively.

The wire–wire transversal interaction between adjacent wires in the $CuO_2$ plane is weak due to the chosen large distance (6a) along x. The wire–wire interaction along z is weak between the $CuO_2$ planes in this layered cuprate perovskite. Correlation is not expected to be an issue for cuprates with doping well away from half-filling of the d-band in agreement with the angular correlation of positron annihilation radiation (ACAR) spectroscopy [59], which probes FSs and bands that evolve with doping in agreement with density-functional theory (DFT) calculations in the overdoped regime.

|  | Cu | $O_{pl}$ | $O_{ap}$ | Total |
|---|---|---|---|---|
| $Hg_{12}Ba_{24}Cu_{12}O_{48}$ | 0.73 | 0.12 | 0.0 | 0.85 |
| $Hg_{12}Ba_{24}Cu_{12}O_{48+2}$ *near O-i stripes* | 1.0 | 0.21 | 0.50 | 1.71 |
| *far from O-i stripes* | 0.93 | 0.18 | 0.15 | 1.26 |

**Table 1.** Calculated local decomposition of the DOS at $E_F$ within the $CuO_6$ tetrahedra (in units of (atom eV)$^{-1}$) for undoped $Hg_{12}Ba_{24}Cu_{12}O_{48}$ and doped $Hg_{12}Ba_{24}Cu_{12}O_{48+n}$ (where n denotes oxygen dopant at O(3) site from ref. [60]. Planar oxygen ions in the O(1) sites of $CuO_2$ plane and apical oxygen ions in the O(2) site are indicated as Opl and Oap, respectively. Total DOS is here for the $CuO_6$ tetrahedron only. The doped oxygen interstitials (O-i) at the O(3) site in the HgO layer form one-dimensional atomic wires. In the doped crystal, the charge density in $CuO(1)_4O(2)_2$ units creates metallic $[CuO_6]^*$ wires that run in the 010 direction. Charge density inhomogeneity is indicated by the peak of the local DOS of 1.76 atom eV$^{-1}$ near the atomic O-i stripes. It appears as if the Fermi level is near a van Hove singularity (VHS) of the local electronic band structure in the overdoped stripes, while the local DOS is 30–35% lower within the separate underdoped portions far from O-i atomic stripes.

## 3. Discussion

We have used scanning nano-X-ray diffraction with a focused beam of 100 × 300 nm$^2$ for directly imaging the mesoscale texture formed by the oxygen interstitial stripes in HgBa$_2$CuO$_{4+y}$. The O-i stripes show a scale-free-like organization, as found at microscopic and atomic scales in scanning micro-X-ray diffraction and previous scanning transmission microscopy measurements in different systems. These results demonstrate the self-similarity in the spatial texture of the oxygen interstitial stripes affecting the material properties. This specific ordering of defects changes dramatically the electronic structure with the rise in a sharp peak at the Fermi energy in the density of state. This indicates that the ordering of interstitial oxygen dopants in atomic stripes gives a metallic character to the CuO$_2$ plane. Therefore, the positive effect on T$_C$ from the ordering of O-i might come from the hole doping that the excess O-i provides to the regions far from the dopants. This scenario is consistent with the recent discovery that the space far from the oxygen interstitial atomic stripes in Hg1201 hosts segregated CDW ordered puddles [57]. The compelling evidence of scale invariant distribution of metallic wires in optimum doping Hg1201 supports the proposal that the system showing the superconducting high-critical temperature is in the proximity of a critical point for nanoscale phase separation near a Lifshitz transition in multi-band correlated electronic materials [64–67] in diborides [68], cuprates [69], organics [70,71], pressurized hydrides [72,73] showing high-order van Hove singularity [74,75], which could be further amplified by spin orbit coupling [76] due to internal electric field gradients driven by local charge segregation in oxygen-doped Hg1201. The compelling evidence of the nematic phase with scale-free distribution of atomic wires in mercury-based superconducting perovskites at optimum doping indicate the presence of critical opalescence [77] of atomic wires at a critical point involving local charge, spin [78,79], and electron-phonon interactions in two-dimensional superconductors [11,80]. A quantum critical point connected with charge density waves has been predicted by Castellani and Arpaia [81,82].

## 4. Materials and Methods

High-energy X-ray diffraction. The O-i organization in Hg1201 has been investigated by high-energy X-ray diffraction (XRD) in transmission geometry at the BW5 beamline of the DESY Synchrotron in Hamburg, using 200 × 200 μm X-ray beam with 100 KeV from the source by a double-crystal Si(111) monochromator. The high-energy allowed us to probe the average bulk O-i organization. The beam line is equipped with a single-axis diffractometer with a motorized goniometric xyz stage head. The single crystal c-axis has been oriented parallel to the direction of the X-ray incoming beam. In this geometry, we can probe the lattice fluctuations on the ab plane. The diffraction patterns have been collected by an area detector at room temperature. We achieved evidence of diffuse streaks connecting Bragg peaks associated with oxygen interstitial (O-i) atomic stripes.

Scanning micro-X-ray diffraction (SµXRD) experiments were performed in reflection geometry using the ID13 beamline at ESRF, Grenoble, France. We applied an incident X-ray energy of 13 KeV. By moving the sample under a 1-µm focused beam with an x–y translator, we scanned a sample area of 65 × 80 µm$^2$, collecting 5200 different diffraction patterns at T = 100 K. For each scanned point of the sample, the intensity profile of the streak from (1,5) to (1,6) Bragg peaks in the hl plane was extracted.

In conclusion, the scanning nano-X-ray diffraction (SnXRD) experimental setup was equipped with a double-crystal monochromator and a Kirkpatrick–Baez mirror system supplied with a beam size of 100 × 300 nm$^2$. The data have been recorded using a wavelength of 0.1 nm. A charge-coupled device FreLon area X-ray detector has been used for the experiment. The sample was mounted on a motorized goniometric xyz stage head. In this study, we achieved evidence of diffuse streaks connecting Bragg peaks associated with oxygen interstitial (O-i) atomic stripes. The diffraction patterns have been collected at room temperature and the intensity profile of the streak from (1,5) to (1,6) Bragg peaks was extracted. We scanned a sample area of 9 × 12 µm$^2$ moving the sample by means of a piezo yz stage with a step size of 100 and 300 nm along the vertical and horizontal directions, respectively.

**Author Contributions:** Conceptualization, G.C. and A.B.; methodology, A.B.; software, G.C. and M.V.M.; validation, G.C., T.J., and A.B.; resources, G.C. and A.B.; data curation, G.C. and M.V.M.; writing—original draft preparation, A.B., G.C., T.J., and M.V.M.; writing—review and editing, G.C., M.V.M. and A.B.; funding acquisition, A.B. All authors have read and agreed to the published version of the manuscript.

**Funding:** M.V.M. acknowledge the Superstripes-onlus for the RICMASS research fellowship.

**Data Availability Statement:** The data that support the findings of this study are available from the corresponding author, G.C., upon reasonable request.

**Acknowledgments:** We thank the 1D13 beamline staff, Manfred Burghammer, Alessandro Ricci, and Nicola Poccia for the experimental help and we are grateful to D. Zhigadlo, S. M. Kazakov for the crystal synthesis.

**Conflicts of Interest:** The authors declare no conflict of interest. The funders had no role in the design of the study; in the collection, analyses, or interpretation of data; in the writing of the manuscript; or in the decision to publish the results.

**Competing Interest Statement:** The authors declare no competing financial interest

**Author Contributions:** Conceptualization, G.C. and A.B. ; methodology, A.B.; software, G.C.; validation, G.C., T.J. and A.B.;, X.X.; resources, X.X.; data curation, X.X.; writing—original draft preparation, A.B., G.C., T.J. M.V.M.; writing—review and editing, G.C., A.B., funding acquisition, A.B. All authors have read and agreed to the published version of the manuscript.

**Funding:** "This research received funding from Superstripes onlus ".

**Data Availability Statement:** The *data* that support the findings of this study are *available* from the corresponding author, [G.C.], *upon reasonable request*.

**Acknowledgments:** We thank the 1D13 beamline staff, Manfred Burghammer, Alessandro Ricci and Nicola Poccia for the experimental help and we are grateful to D. Zhigadlo, S. M. Kazakov for the crystal synthesis.

**Conflicts of Interest:** The authors declare no conflict of interest. The funders had no role in the design of the study; in the collection, analyses, or interpretation of data; in the writing of the manuscript; or in the decision to publish the results.

## References

1. Bussmann-Holder, A. Keller, H. and Bianconi A., eds., High-T$_c$ copper oxide superconductors and related novel materials (Springer Series in Materials Science; Springer International Publishing AG, Vol 255, **2017**
2. Egami, T., Alex and the origin of high-temperature superconductivity. In *High-Tc Copper Oxide Superconductors and Related Novel Materials* **2017**, (pp. 35-46). Springer, Cham
3. Giraldo-Gallo, P., Zhang, Y., Parra, C., Manoharan, H. C., Beasley, M. R., Geballe, T. H., et al. Stripe-like nanoscale structural phase separation in superconducting $BaPb_{1-x}Bi_xO_3$. *Nature* **2015**, *Communications*, *6*(1), 1-9.
4. Krockenberger, Y., Ikeda, A., Yamamoto, H. Atomic Stripe Formation in Infinite-Layer Cuprates. *ACS omega*, **2021**, *6*(34), 21884-21891.
5. Bartolomé, E., Mundet, B., Guzmán, R., Gázquez, J., Valvidares, S. M., Herrero-Martín, J., et al. Embedded magnetism in $YBa_2Cu_3O_7$ associated with Cu–O vacancies within nanoscale intergrowths: implications for superconducting current performance. *ACS Applied Nano Materials* **2020**, *3*(3), 3050-3059 https://doi.org/10.1021/acsanm.0c00505
6. Gavrichkov, V. A., Shan'ko, Y., Zamkova, N. G., and Bianconi, A. Is there any hidden symmetry in the stripe structure of perovskite high-temperature superconductors? *The Journal of Physical Chemistry Letters*, **2019**, *10* 8), 1840-1844
7. Bianconi, A. Shape resonances in superstripes. *Nature Physics*, **2013,** *9*(9), 536-53
8. Hoshi, K., Sakuragi, S., Yajima, T., Goto, Y., Miura, A., Moriyoshi, C., et al. Structural phase diagram of $LaO_{1-x}F_xBiSSe$: suppression of the structural phase transition by partial F substitutions. *Condensed Matter*, **2020,** *5*(4), 81.
9. Athauda, A., and Louca, D. Nanoscale atomic distortions in the $BiS_2$ superconductors: ferrodistortive sulfur modes. *Journal of the Physical Society of Japan*, **2019**, *88*(4), 041004.

10. Li, W. M., Zhao, J. F., Cao, L. P., Hu, Z., Huang, Q. Z., Wang, X. C., et al. Superconductivity in a unique type of copper oxide. *Proceedings of the National Academy of Sciences*, **2019,** *116*(25), 12156-12160.
11. Conradson, S. D., Geballe, T. H., Gauzzi, A., Karppinen, M., Jin, C., Baldinozzi, G., et al. Local lattice distortions and dynamics in extremely overdoped superconducting $YSr_2Cu_{2.75}Mo_{0.25}O_{7.54}$. *Proceedings of the National Academy of Sciences*, **2020,** 117(9), 4559-4564.
12. Crabtree, G. W., Sarrao, J. L. Opportunities for mesoscale science. *MRS Bulletin* **2012**, 37, 1079-1088
13. Schroer, C. G. et al. Hard X-ray nanoprobe based on refractive X-ray lenses. *Applied Physics Letters* **2005,** 87, 124103.
14. Campi, G. Structural Fluctuations at Nanoscale in Complex Functional Materials. In Synchrotron Radiation Science and Applications **2021**, (pp. 181-189). Springer, Cham.
15. Campi, G., and Bianconi, A. Evolution of complexity in out-of-equilibrium systems by time-resolved or space-resolved synchrotron radiation techniques. *Condensed Matter*, **2019,** *4*(1), 32.
16. Phillabaum, B., Carlson, E. W. and Dahmen, K. A. Spatial complexity due to bulk electronic nematicity in a superconducting underdoped cuprate. *Nat. Commun.* **2012**, 3, 915.
17. Carlson, E. W., Liu, S., Phillabaum, B., and Dahmen, K. A. Decoding spatial complexity in strongly correlated electronic systems. *Journal of Superconductivity and Novel Magnetism*, **2015**, *28*(4), 1237-1243.
18. Liu, S., Carlson, E. W., and Dahmen, K. A. Connecting complex electronic pattern formation to critical exponents. *Condens. Matter* **2021**, *6*(4), 39; https://doi.org/10.3390/condmat6040039
19. Li, J., Pelliciari, J., Mazzoli, C., Catalano, S., Simmons, F., Sadowski, J. T., Levitan, A., Gibert, M., Carlson, E., Triscone J.-M. et al. Scale-invariant magnetic textures in the strongly correlated oxide $NdNiO_3$. *Nat Commun* **2019**, 10, 4568 https://doi.org/10.1038/s41467-019-12502-0
20. Campi, G., Poccia, N., Joseph, B., Bianconi, A., Mishra, S., Lee, J., et al. Direct visualization of spatial inhomogeneity of spin stripes order in $La_{1.72}Sr_{0.28}NiO_4$. *Condensed Matter*, **2019**, *4*(3), 77.
21. Campi, G.; Bianconi, A.; Ricci, A. Nanoscale phase separation of incommensurate and quasi-commensurate spin stripes in low temperature spin glass of $La_{2-x}Sr_xNiO_4$. *Condensed Matter* **2021**, 6, 45. https://doi.org/10.3390/condmat6040045
22. Richardella, A. et al. Visualizing critical correlations near the metal-insulator transition in $Ga_{1-x}Mn_xAs$. *Science* 2**010,** 327, 665–669.
23. Newman, M. E., Barabási, A. L. E., & Watts, D. J. (2006). The structure and dynamics of networks. Princeton university press..
24. Richard, P., Valance, A., Métayer, J. F., Sanchez, P., Crassous, J., Louge, M., & Delannay, R. (2008). Rheology of confined granular flows: Scale invariance, glass transition, and friction weakening. Physical review letters, 101(24), 248002.

25. Bianconi, G., Enhancement of $T_c$ in the superconductor–insulator phase transition on scale-free networks. *Journal of notal Mechanics: Theory and Experiment*, **2012**, *2012*(07), P07021.
26. Uchino, T., Teramachi, N., Matsuzaki, R., Tsushima, E., Fujii, S., Seto, Y., et al. Proximity coupling of superconducting nanograins with fractal distributions. *Physical Review B*, **2020**, *101*(3), 035146.
27. Bianconi, A. Shape resonances in superconducting gaps and complexity in superstripes. *J Supercond Nov Magn* **2013**, 26**,** 2821–2827 https://doi.org/10.1007/s10948-013-2205-5
28. Ashkenazi, J., and Johnson, N. F. Quantum criticality stabilizes high $T_c$ superconductivity against competing symmetry-breaking instabilities. *Journal of superconductivity and novel magnetism*, **2013**, *26*(8), 2611-2616.
29. Campi, G., and Bianconi, A. High-Temperature superconductivity in a hyperbolic geometry of complex matter from nanoscale to mesoscopic scale. *Journal of Superconductivity and Novel Magnetism*, **2016,** *29*(3), 627-631.
30. Littlewood, P. Superconductivity: an X-ray oxygen regulator. *Nature Materials* **2011**, 10, 726-727.
31. Perrichon, A. et al. Lattice dynamics modified by excess oxygen in $Nd_2NiO_{4+\delta}$: Triggering Low-Temperature oxygen diffusion. *J. Phys. Chem. C* **2015**, 119, 1557-1564. DOI: 10.1021/jp510392h.
32. Campi, G., Ricci, A., Poccia, N., Fratini, M., Bianconi, A. X-Rays Writing/Reading of charge density waves in the $CuO_2$ plane of a simple cuprate superconductor. *Condensed Matter*, **2017**, *2*(3), 26.
33. Bianconi, A., Marcelli, A., Bendele, M., Innocenti, D., Barinov, A., Poirot, N., and Campi, G. VUV Pump and probe of phase separation and oxygen interstitials in $La_2NiO_{4+y}$ using spectromicroscopy. *Condensed Matter*, **2018,** *3*(1), 6
34. Jarlborg, T., Bianconi, A. Electronic structure of superoxygenated $La_2NiO_4$ domains with ordered oxygen interstitials. *Journal of Superconductivity and Novel Magnetism*, **2016,** 29 (3), 615-621
35. Fratini, M. et al. Scale-free structural organization of oxygen interstitials in $La_2CuO_{4+y}$. *Nature* **2010,** 466, 841-844
36. Poccia, N. et al. Evolution and control of oxygen order in a cuprate superconductor. *Nature Materials* **2011**, 10, 733-736 (2011).
37. Poccia, N. et al. Optimum inhomogeneity of local lattice distortions in $La_2CuO_{4+y}$. *Proceedings of the National Academy of Sciences* **2012**, 109, 15685-15690
38. Poccia, N., Bianconi, A., Campi, G., Fratini, M. and Ricci, A. Size evolution of the oxygen interstitial nanowires in $La_2CuO_{4+y}$ by thermal treatments and X-ray continuous illumination. *Superconductor Science and Technology* **2012**, 25, 124004
39. Poccia, N. et al. Percolative superconductivity in $La_2CuO_{4.06}$ by lattice granularity patterns with scanning micro X-ray absorption near edge structure. *Applied Physics Letters* **2014**, 104, 221903.
40. Lee, K. H., and Hoffmann, R. Oxygen interstitials in superconducting $La_2CuO_4$: their valence state and role. *The Journal of Physical Chemistry A,* **2006,** 110(2), 609–617. doi:10.1021/jp053154f

41. De Mello, E. V. L. Describing how the superconducting transition in $La_2CuO_{4+y}$ is related to the i-O phase separation. *Journal of Superconductivity and Novel Magnetism* **2012,** 25, 1347-1350.
42. Jarlborg, T., and Bianconi, A. Fermi surface reconstruction of superoxygenated $La_2CuO_4$ superconductors with ordered oxygen interstitials. *Physical Review B*, **2013,** 87 (5), 054514.
43. Ricci, A. et al. Multiscale distribution of oxygen puddles in 1/8 doped $YBa_2Cu_3O_{6.67}$. *Scientific Reports* **2013**, *3,* 2383.
44. Ricci, A. et al. Networks of superconducting nano-puddles in 1/8 doped $YBa_2Cu_3O_{6.5+y}$ controlled by thermal manipulation. *New Journal of Physics* **2014,** 16, 053030.
45. Campi, G. et al. Scanning micro-X-ray diffraction unveils the distribution of oxygen chain nanoscale puddles in $YBa_2Cu_3O_{6.33}$. *Physical Review B* **2013**, 87, 014517.
46. Bianconi, A., Lusignoli, M., Saini, N. L., Bordet, P., Kvick, Å., Radaelli, P. G. Stripe structure of the $CuO_2$ plane in $Bi_2Sr_2CaCu_2O_{8+y}$ by anomalous X-ray diffraction. *Physical Review B*, **1996,** *54*(6), 4310.
47. Poccia, N. et al. Spatial inhomogeneity and planar symmetry breaking of the lattice incommensurate supermodulation in the high-temperature superconductor $Bi_2Sr_2CaCu_2O_{8+y}$. *Physical Review B* **2011**, 84, 100504.
48. Zeljkovic, I., Xu, Z., Wen, J., Gu, G., Markiewicz, R. S., and Hoffman, J. E. Imaging the impact of single oxygen atoms on superconducting $Bi_2Sr_{2-y}CaCu_2O_{8+x}$. *Science*, **2012**, *337*(6092), 320-323.
49. Poccia, N., Zhao, S. Y. F., Yoo, H., Huang, X., Yan, H., Chu, Y. S., et al. Spatially correlated incommensurate lattice modulations in an atomically thin high-temperature $Bi_{2.1}Sr_{1.9}CaCu_{2.0}O_{8+y}$ superconductor. *Phys. Rev. Materials*, **2020,** *4*(11), 114007.
50. Putilin, S., Antipov, E., Chmaissem, O. et al. Superconductivity at 94 K in $HgBa_2CuO_{4+\delta}$. *Nature* **1993**, 362, 226–228. https://doi.org/10.1038/362226a0
51. Karpinski, J. et al. High-pressure synthesis, crystal growth, phase diagrams, structural and magnetic properties of $Y_2Ba_4Cu_nO_{2n+x}$, $HgBa_2Ca_{n-1}Cu_nO_{2n+2+x}$ and quasi-one-dimensional cuprates. *Supercond. Sci. Technol*. **1999**, 12, R153–R181. https://doi.org/10.1088/0953-2048/12/9/201
52. Legros, A., Loret, B., Forget, A., Bonnaillie, P., Collin, G., Thuéry, P., et al. Crystal growth and doping control of $HgBa_2CuO_{4+\delta}$, the model compound for high-$T_c$ superconductors. *Materials Research Bulletin*, **2019**, *118*, 110479.
53. Bordet, P., Duc, F., LeFloch, S., Capponi, J. J., Alexandre, E., Rosa-Nunes, M., et al. Single crystal X-ray diffraction study of the $HgBa_2CuO_{4+\delta}$ superconducting compound. *Physica C: Superconductivity*, **1996,** *271*(3-4), 189-196.
54. Balagurov, A. M., Sheptyakov, D. V., Aksenov, V. L., Antipov, E. V., Putilin, S. N., Radaelli, P. G., and Marezio, M. Structure of $HgBa_2CuO_{4+\delta}$ (0.06<δ<0.19) at ambient and high pressure. *Phys. Rev. B*, **1999**, 59(10), 7209–7215. doi:10.1103/physrevb.59.7209
55. Auvray, N., Loret, B., Chibani, S., Grasset, R., Guarnelli, Y., Parisiades, P., ... Sacuto, A. Exploration of Hg-based cuprate superconductors by Raman spectroscopy under hydrostatic pressure. *Phys. Rev. B,* **2021**, 103(19), 19513 doi:10.1103/physrevb.103.195130

56. Izquierdo, M. et al. One dimensional ordering of doping oxygen in superconductors evidenced by X-ray diffuse scattering. *Journal of Physics and Chemistry of Solids* **2011**, 72, 545-548.
57. Campi, G., Bianconi, A., Poccia, N. et al. Inhomogeneity of charge-density-wave order and quenched disorder in a high-$T_c$ superconductor. *Nature* **2015**, 525, 359-362. https://doi.org/10.1038/nature14987
58. Izquierdo, M., Freitas, D. C., Colson, D., Garbarino, G., Forget, A., Raffy, H., Itié, J-P., Ravy, S., Fertey, P. and Núñez-Regueiro, M. Charge order and suppression of superconductivity in $HgBa_2CuO_{4+d}$ at high pressures. *Condensed Matter*, **2021,** 6(3), 2.
59. Barbiellini, B., and Jarlborg, T. Electron and positron states in $HgBa_2CuO_4$. *Physical Review B*, **1994,** *50*(5), 3239.
60. Jarlborg, T., Bianconi, A. Electronic structure of $HgBa_2CuO_{4+\delta}$ with self-organized interstitial oxygen wires in the Hg spacer planes. *Journal of Superconductivity and Novel Magnetism*, **2018**, *31*(3), 689-695.
61. Wu, J., Bollinger, A. T., He, X., & Božović, I. (2017). Spontaneous breaking of rotational symmetry in copper oxide superconductors. Nature, 547(7664), 432-435.
62. Wahlberg, E., Arpaia, R., Seibold, G., Rossi, M., Fumagalli, R., Trabaldo, E., Brookes N. B., Braicovich L., Caprara S., Gran U., Ghiringhelli G., Bauch T. & Lombardi, F. (2021). Restored strange metal phase through suppression of charge density waves in underdoped YBa2Cu3O7–δ. Science, 373(6562), 1506-1510.
63. Metropolis N., Rosenbluth A.W., Rosenbluth M.N., Teller A.H., Teller E.J., Equation of state calculations by fast computing machines, *J. Chem. Phys* **1953**, 21, 1087.
64. Proffen Th., Neder R.B., DISCUS, a program for diffuse scattering and defect structure simulations, *J. Appl. Cryst.* **1997,** 30, 171.
65. Bianconi, A., Poccia, N., Sboychakov, A. O., Rakhmanov, A. L., and Kugel, K. I. Intrinsic arrested nanoscale phase separation near a topological Lifshitz transition in strongly correlated two-band metals. *Superconductor Science and Technology*, **2015**, *28*(2), 024005.
66. Jaouen, T., Hildebrand, B., Mottas, M. L., Di Giovannantonio, M., Ruffieux, P., Rumo, M., Nicholson, C. W., Razzoli, E., Barreteau, C., Ubaldini, A., Giannini, E., Vanini, F., Beck, H., Monney, C. and Aebi, P. Phase separation in the vicinity of Fermi surface hot spots. *Physical Review B*, **2019**, *100*(7), 075152. doi:10.1103/PhysRevB.100.075152
67. Jarlborg, T., Bianconi, A. Multiple electronic components and Lifshitz transitions by oxygen wires formation in layered cuprates and nickelates. *Condensed Matter*, **2019,** 4 (1), 15.
68. Agrestini, S., Metallo, C., Filippi, M., Simonelli, L., Campi, G., Sanipoli, C., Liarokapis, E., De Negri, S., Giovannini, M., Saccone, A., Latini, A. and Bianconi, A. Substitution of Sc for Mg in $MgB_2$: Effects on transition temperature and Kohn anomaly. *Physical Review B*, **2004,** *70*(13), 134514.
69. Perali, A., Innocenti, D., Valletta, A., and Bianconi, A. Anomalous isotope effect near a 2.5 Lifshitz transition in a multi-band multi-condensate superconductor made of a superlattice of stripes. *Superconductor Science and Technology*, **2012,** *25*(12), 124002.

70. Mazziotti, M. V., Valletta, A., Campi, G., Innocenti, D., Perali, A., and Bianconi, A. Possible Fano resonance for high-Tc multi-gap superconductivity in p-Terphenyl doped by K at the Lifshitz transition. *EPL (Europhysics Letters)*, **2017**, 118 (3), 37003.
71. Pinto, N., Di Nicola, C., Trapananti, A., Minicucci, M., Di Cicco, A., Marcelli, A., Bianconi, A., Marchetti, F. and Pettinari, C., and Perali, A. (2020). Potassium-doped para-terphenyl: Structure, electrical transport properties and possible signatures of a superconducting transition. *Condensed Matter*, *5*(4), 78.
72. Bianconi, A., and Jarlborg, T. Lifshitz transitions and zero point lattice fluctuations in sulfur hydride showing near room temperature superconductivity. *Novel Superconducting Materials*, 2015, *1*(1), 37-49 https://doi.org/10.1515/nsm-2015-0006
73. Mazziotti, M. V., Jarlborg, T., Bianconi, A., and Valletta, A. Room temperature superconductivity dome at a Fano resonance in superlattices of wires. *EPL (Europhysics Letters)*, **2021**, 134 (1), 17001
74. Mazziotti, M.V. et al. Resonant multi-gap superconductivity at room temperature near a Lifshitz topological transition in sulfur hydrides, Jour*nal of Applied Physics* **2021,** 130, 173904 https://doi.org/10.1063/5.0070875
75. H. Isobe and L. Fu, Supermetal, *Phys. Rev. Res*. **2019**, 1, 033206. https://doi.org/10.1103/PhysRevResearch.1.033206
76. Mazziotti, M. V., Valletta, A., Raimondi, R., Bianconi, A. Multigap superconductivity at an unconventional Lifshitz transition in a three-dimensional Rashba heterostructure at the atomic limit. *Physical Review B*, **2021,** 103 (2), 02452.
77. Walsh, C., Sémon, P., Sordi, G., & Tremblay, A. M. (2019). Critical opalescence across the doping-driven Mott transition in optical lattices of ultracold atoms. Physical Review B, 99(16), 165151.
78. Di Castro, D., Bianconi, G., Colapietro, M., Pifferi, A., Saini, N. L., Agrestini, S., Bianconi, A. Evidence for the strain critical point in high Tc superconductors. *The European Physical Journal B-Condensed Matter and Complex Systems*, **2000**, *18*(4), 617-624.
79. Gavrichkov, V. A., Polukeev, S. I. Magnetic interaction in doped 2D perovskite cuprates with banoscale inhomogeneity: lattice nonlocal effects vs superexchange. *arXiv preprint* **2022.** *arXiv:2205.11959*.
80. Benedek, G., Manson, J. R., Miret-Artés, S., Ruckhofer, A., Ernst, W. E., Tamtögl, A., and Toennies, J. P. Measuring the electron–phonon interaction in two-dimensional superconductors with he-atom scattering. *Condensed Matter*, **2020**, *5*(04), 79.
81. Arpaia, R., Caprara, S., Fumagalli, R., De Vecchi, G., Peng, Y. Y., Andersson, E., Betto, D., De Luca, G.M., Brookes, N.B., Lombardi, F., Braicovich, L., Di Castro, C., Grilli, M., Salluzzo, M. & Ghiringhelli, G. (2019). Dynamical charge density fluctuations pervading the phase diagram of a Cu-based high-T c superconductor. Science, 365(6456), 906-910.
82. Castellani, C., Di Castro, C., & Grilli, M. (1995). Singular quasiparticle scattering in the proximity of charge instabilities. Physical review letters, 75(25), 4650.